# On a Generalized Foster-Lyapunov Type Criterion for the Stability of Multidimensional Markov chains with Applications to the Slotted-Aloha Protocol with Finite Number of Queues [*]

Sayee C. Kompalli and Ravi R. Mazumdar
Department of Electrical and Computer Engineering
University of Waterloo, Canada N2L 3G1.
Email: skompall@uwaterloo.ca, mazum@ece.uwaterloo.ca

June 5, 2018

## Abstract

In this paper, we generalize a positive recurrence criterion for multidimensional discrete-time Markov chains over countable state spaces due to Rosberg (JAP, Vol. 17, No. 3, 1980). We revisit the stability analysis of well known slotted-Aloha protocol with finite number of queues. Under standard modeling assumptions, we derive a sufficient condition for the stability by applying our positive recurrence criterion. Our sufficiency condition for stability is linear in arrival rates and does not require knowledge of the stationary joint statistics of queue lengths. We believe that the technique reported here could be useful in analyzing other stability problems in countable space Markovian settings. Toward the end, we derive some sufficient conditions for instability of the protocol.

## 1 Introduction

Recently there has been much activity in trying to understand the stability region of different multipleaccess schemes for wireless networks of which the Aloha protocol and its variation Slotted-Aloha protocol are archetypes. The most common access scheme for wireless networks is the contention mechanism used in the IEEE 802.11 that uses a exponential window type of backoff where the window doubles if collisions occur. Due to the difficulty of analyzing the stability region much attention has been focussed on the so called saturation throughput. This is the average throughput seen by each queue or user assuming that they always have a packet to transmit. Thus the entire question of stability is sidestepped. It is well known and by now there are many text-book accounts of it [1] that the slotted aloha scheme is unstable when the number of users (or queues) goes to infinity and there is a fixed probability of attempting to transmit. It is only the finite case that is really unknown.

---

[*]Preliminary version of the results reported here are to appear in the Proceedings of 21st International Teletraffic Congress (ITC 21), Paris.

The stability analysis of slotted Aloha protocol with finite number of queues has attracted lot of attention from researchers since its formulation by Tsybakov and Mikhailov [2] in 1979. The continued interest in this protocol is chiefly due to the fact that in spite of its extreme simplicity the analytical difficulties presented by the interacting queues has yielded no general necessary and sufficient conditions for positive recurrence. The standard modeling assumptions made in the literature to analyze this protocol result in a discrete-time Markov chain model of the queue lengths over a countable set. An important performance measure of this protocol is stability, i.e., for what set of arrival rates at different queues the average delay experienced by a packet is finite. In this paper, we resurrect a positive recurrence criterion for countable space multidimensional Markov chain proved by Rosberg [3] that has been largely forgotten. We show usefulness of this criterion by first generalizing criterion [3] due to Rosberg and then applying it to the stability analysis of buffered slotted-Aloha protocol with finite number of queues.

The discrete-time countable space Markov chain modeling of the slotted-Aloha protocol with finite number of queues was first proposed and analyzed by Tsybakov and Mikhailov in [2]. In [2], they provided the exact stability characterization when the number of queues $J = 2$. The next effort was by Rao and Ephremides [4] in 1988 who provided *exact* stability conditions for $J = 2$, and sufficient conditions for stability for $J > 2$ using stochastic dominance arguments and assuming Bernoulli input process at each queue. In 1994, Szpankowski [5] obtained the exact stability region for $J > 2$. The stability region characterization is given in terms of the stationary probabilities of joint statistics of the queues. However, for systems with more than three queues, the necessary and sufficient condition cannot be computed explicitly since it becomes very hard to compute the joint stationary statistics of the queues. In 1999, Luo and Ephremides [6] introduced the concept of "stability rank" to obtain tight inner and outer bounds to the stability region when $J > 2$. When queue $i$ is known to be stable, then any queue $j$ such that $\frac{\lambda_j(1-p_j)}{p_j} \leq \frac{\lambda_i(1-p_i)}{p_i}$ is also proven to be stable, where $\lambda_j$ and $p_j$ are, respectively, the average packet arrival rate and the fixed transmission attempt probability of the $j$th queue. Then it immediately follows that if queue $k$ is unstable then $\frac{\lambda_k(1-p_k)}{p_k} > \frac{\lambda_i(1-p_i)}{p_i}$. With the help of stability ranks, they computed tight inner bounds to the stability region. Unfortunately, here also it is required to determine some stationary joint probabilities but which are extremely difficult to compute.

In all these papers, the goal was to derive sufficiency conditions for a *fixed* transmission attempt probability vector $p$. Instead, if one considers the union of the stability regions over all possible transmission probability vectors $p$, one obtains the closure of the stability region. In 1991, Anantharam [7] obtained closure of the stability region for $J > 2$ albeit for a correlated arrival processes.

Recently, in [8], a simple approximate expression for the stability region was proposed by using *mean field analysis*. Indeed they show a propagation of chaos takes place when the number of interacting queues is large. This expression is proved to be exact when the number of queues grows large, and is also shown to be accurate in the case of small-queue systems through numerical experiments. The approximate stability region is derived assuming that queue lengths evolve independently. The boundary of the approximate stability region is characterized by a parametric expression that is a function of the attempt probability vector $p$. Our sufficient conditions for stability are in the form of simple linear inequalities and hence lead to much easier verification.

Our approach leads to sufficient conditions that do not depend on knowing the stationary distributions, and are completely characterized by the arrival parameters and the attempt probabilities of the queues. In particular, we show that for the case of two and three interacting queues we can recover the known results. However, we have found a few instances where certain arrival rate vectors $\lambda$ satisfy the sufficiency condition for stability derived by Luo and Epphremides [6] but *not* ours. However we do not need to establish the stability of any higher rank queue as they require.

A popular technique to establish ergodicity of countable space Markov chains is through Foster-Lyapunov approach which consists of finding a function that satisfies Foster's criterion [9]. Though this approach proved to be very successful for one dimensional Markov chains, but finding such a function proved to be very difficult in the case of multidimensional Markov chains. Rosberg [3] extended Foster's criterion [9] for $J$-dimensional Markov chains by requiring existence of $J$ functions, one for each coordinate of the process. But the applicability of his criterion became limited by the difficulties that arise in verifying his conditions. The main contribution of this paper is in proposing a generalization of the positive recurrence criterion due to Rosberg [3], thus expanding the scope of its applicability, and also distilling some of the assumptions made therein into a form that immediately lead to easier verification. To illustrate the applicability of our criterion, we derive sufficient conditions for stability of the well known slotted-Aloha protocol for multiaccess communication.

The rest of the paper is organized into two major parts: Section 2 and Section 3. In Section 2, we generalize the positive recurrence criterion for multidimensional countable space Markov chains due to Rosberg [3]. In Section 3, we apply the positive recurrence criterion developed in Section 2 to the stability analysis of the slotted-Aloha protocol with finite number of queues. We provide some remarks on the instability of the protocol in Section 3.4. We end the paper with conclusions in Section 4.

## 2 A Generalized Foster-Lyapunov Type Criterion for Positive Recurrence

Let $\mathcal{X}$ be a countable set of states over which the irreducible, aperiodic, and discrete-time Markov chain $\{X^n, n \geq 0\}$ takes its values. For any integer $k \geq 1$, define $\{p_{xy}^k, x, y \in \mathcal{X}\}$ to be the $k$-step transition probability law of the Markov chain $\{X^n, n \geq 0\}$. For any subset $\mathcal{B} \subseteq \mathcal{X}$, we know that the $\lim_{k \to \infty} p_{x\mathcal{B}}^k = \lim_{k \to \infty} \sum_{y \in \mathcal{B}} p_{xy}^k = \pi(\mathcal{B}) \geq 0$ exists and is independent of the initial state $x$. For any nonnegative-valued function $V$ on $\mathcal{X}$, let us define $\Delta^k V(x) \triangleq \sum_y p_{xy}^k V(y) - V(x)$ to be the $k$-step drift of the function $V$ in state $x$. Let us now define the notion of "partial order" on the state space $\mathcal{X}$. Let $\preceq$ be a binary relation on the set $\mathcal{X}$ such that

(i) $x \preceq x$ for all $x \in \mathcal{X}$ (reflexivity)

(ii) $x \preceq y$ and $y \preceq z$ imply $x \preceq z$ (transitivity)

(iii) $x \preceq y$ and $y \preceq x$ imply $x = y$ (antisymmetry)

Then $\preceq$ is called a *partial order* on the set $\mathcal{X}$. An element $x^* \in \mathcal{X}$ is called the *minimal* element of $\mathcal{X}$ with respect to the partial order $\preceq$ if $x^* \preceq y$ for all $y \in \mathcal{X}$. Let us now suppose that $\preceq$ defines a partial order on the set $\mathcal{X}$, and the Markov chain evolution over the discrete-time $n$ is modeled by the stochastic mapping $f$ defined as

$$X^{n+1} = f(X^n, \Lambda^n), \quad n \geq 0 \qquad (1)$$

where $X^n \in \mathcal{X}$ and $\Lambda^n$ is the input (driving function) to the system. We will see two instances of this stochastic recursive relation in equations (6) and (7) in the context of slotted-Aloha protocol. Let us say that the mapping $f$ is *order-preserving* when, for a *fixed* input $\Lambda^n$, we have

$$X^n \preceq Y^n \;\Rightarrow\; X^{n+1} = f(X^n, \Lambda^n) \preceq f(Y^n, \Lambda^n) = Y^{n+1} \qquad (2)$$

We now state our main contribution, which is a generalization of the multidimensional positive recurrence criterion due to Rosberg [3], as the following theorem. In the rest of the paper, we will be using the same notation $\mathcal{X}$ to denote both the state space and its subsets. The distinction is made through the usage of subscripts, i.e., $\mathcal{X}_j$ denotes a subset.

**Theorem 2.1** *Let $J \geq 2$ be an integer.*

**Assumption 2.1** *There exists a collection $\mathcal{P} = \{\mathcal{P}_1, \mathcal{P}_2, \ldots, \mathcal{P}_J\}$ of partitions of the set $\mathcal{X}$ where $\mathcal{P}_j = \{\mathcal{X}_j, \mathcal{X}_j^c\}$, and nonnegative-valued functions $\{V_j(x), x \in \mathcal{X}\}$ for $1 \leq j \leq J$ such that the drift $\Delta V_j(x)$ of the function $V_j$ in the state $x$ has the following form:*

$$\Delta V_j(x) \leq \begin{cases} \eta_j & \text{for } x \in \mathcal{X} \\ -\epsilon_j & \text{for } x \in \mathcal{X}_j^c, \end{cases} \qquad (3)$$

*where $\epsilon_j > 0$ and $\eta_j \geq 0$.*

**Assumption 2.2** *There exist partitions $\{\mathcal{A}_{j,k}, \mathcal{A}_{j,k}^c\}$, $k \geq 1$ and $1 \leq j \leq J$, of the set $\mathcal{X}$ with the following two properties:*

(i) $p_{xy}^l = 0$, $0 \leq l \leq k-1$, for $x \in \mathcal{A}_{j,k}^c$ and $y \in \mathcal{X}_j$

(ii) $\cap_j \mathcal{A}_{j,k}$ is a finite set

**Assumption 2.3** *The stochastic recursive relation $f$ (equation (1)) that models the system is order-preserving, and for $1 \leq j \leq J$ the drift function $\Delta V_j(x)$ in the argument $x$ is a non-increasing function with respect to the partial order $\preceq$, i.e., $x \preceq y \Rightarrow \Delta V_j(x) \geq \Delta V_j(y)$.*

*Then the Markov chain $\{X_n, n \geq 1\}$ is positive recurrent.* ∎

Before we formally prove Theorem 2.1 and then point out how Theorem 2.1 generalizes the positive recurrence criterion of Rosberg, we briefly discuss a multivariate stochastic order known as *usual multivariate stochastic order* [10] and also establish few supporting Lemmas.

## 2.1 Multivariate Stochastic Order for Random Variables

Let $\preceq$ be a partial order on the set $\mathcal{X}$. For any two elements $x$ and $y$ in $\mathcal{X}$, we say that a set $\mathcal{B} \subseteq \mathcal{X}$ is (i) an *Upper Set* if $y \in \mathcal{B}$ whenever $x \preceq y$ and $x \in \mathcal{B}$ and (ii) a *Lower Set* if $y \in \mathcal{B}$ whenever $y \preceq x$ and $x \in \mathcal{B}$. For any two random variables $X$ and $Y$ that take values in the set $\mathcal{X}$, we say that the random variable $X$ is *stochastically larger* than the random variable $Y$ if

$$p(X \in \mathcal{B}) \geq p(Y \in \mathcal{B}), \quad \forall \text{ Upper sets } \mathcal{B} \subseteq \mathcal{X}$$

When $X$ is stochastically larger than $Y$, we write $X \geq_{st} Y$. We say that $X$ is stochastically smaller than $Y$, i.e., $X \leq_{st} Y$, if

$$p(X \in \mathcal{B}) \geq p(Y \in \mathcal{B}), \quad \forall \text{ Lower sets } \mathcal{B} \subseteq \mathcal{X}$$

An important characterization of the usual stochastic order is given in the following theorem due to Strassen [11].

**Theorem 2.2 (Theorem 6.B.1. in [10])** *The random vectors $X$ and $Y$ satisfy $X \leq_{st} Y$ if, and only if, there exists two random vectors $\hat{X}$ and $\hat{Y}$, defined on the same probability space, such that $\hat{X} =_{st} X$, $\hat{Y} =_{st} Y$, and $P\left\{\hat{X} \leq \hat{Y}\right\} = 1$.* ∎

In Lemma 2.1, we state an expression for $\Delta^k V(x)$ in terms of the one-step drifts $\{\Delta V(x), x \in \mathcal{X}\}$.

**Lemma 2.1** *Let $t_1$ and $t_2$ be positive integers. Then*

$$\Delta^{t_1+t_2} V(x) = \Delta^{t_1} V(x) + \sum_{y \in \mathcal{X}} p_{xy}^{t_1} \Delta^{t_2} V(y)$$

∎

*Proof:*

$$\begin{aligned}
\Delta^{t_1+t_2} V(x) &= \sum_{y \in \mathcal{X}} p_{xy}^{t_1+t_2} V(y) - V(x) \\
&= \sum_{y \in \mathcal{X}} \left( \sum_{z \in \mathcal{X}} p_{xz}^{t_1} p_{zy}^{t_2} \right) V(y) - V(x) \\
&= \sum_{z \in \mathcal{X}} p_{xz}^{t_1} \sum_{y \in \mathcal{X}} p_{zy}^{t_2} V(y) - V(x) \\
&= \sum_{z \in \mathcal{X}} p_{xz}^{t_1} \left( \Delta^{t_2} V(z) + V(z) \right) - V(x) \\
&= \Delta^{t_1} V(x) + \sum_{y \in \mathcal{X}} p_{xy}^{t_1} \Delta^{t_2} V(y)
\end{aligned}$$

∎

**Corollary 2.1** *Define $t_0 = 0$. For some integer $J \geq 1$, let $t_1, t_2, \ldots, t_J$ be positive integers. Then for $x \in \mathcal{X}$,*

$$\Delta^{\sum_{j=1}^{J} t_j} V(x) = \sum_{y \in \mathcal{X}} \sum_{j=0}^{J-1} p_{xy}^{(\sum_{k=0}^{j} t_k)} \Delta^{t_{j+1}} V(y)$$

∎

*Proof:* Repeated application of Lemma 2.1 gives the result. ∎

Next, we establish that $\lim_{k \to \infty} \frac{1}{k} \Delta^k V(x) = c^* \geq 0$ under the assumption that the Markov chain is irreducible and aperiodic. To see this, assume that the drift $\Delta V(x)$ is upper bounded by a positive constant $\eta$. Then it is easy to see the existence of $\lim_{k \to \infty} \frac{1}{k} \Delta^k V(x)$ because $\lim_{k \to \infty} \frac{1}{k} \Delta^k V(x) = \lim_{k \to \infty} \frac{1}{k} \sum_{y \in \mathcal{X}} \Delta V(y) \sum_{l=0}^{k-1} p_{xy}^l \leq \eta$. Moreover, since $\frac{V(x)}{k} \xrightarrow{k \to \infty} 0$ and $V(x) \geq 0$, we have

$$\lim_{k \to \infty} \frac{1}{k} \Delta^k V(x) = \lim_{k \to \infty} \frac{1}{k} \left( \sum_{y \in \mathcal{X}} p_{xy}^k V(y) - V(x) \right) = c^* \geq 0$$

.

**Lemma 2.2** *For any non-negative random variables $Y_1, Y_2, \ldots, Y_m$ and a constant $a > 0$,*

$$p\left( \max_{1 \leq i \leq m} Y_i \geq a \right) \leq \frac{1}{a} \sum_{i=1}^{m} \mathbb{E}(Y_i)$$

*Proof:*

$$p\left( \max_{1 \leq i \leq m} Y_i \geq a \right) = p\left( \bigcup_{i=1}^{m} \{Y_i \geq a\} \right)$$

$$\leq \sum_{i=1}^{m} p(Y_i \geq a)$$

$$\overset{(b)}{\leq} \frac{1}{a} \sum_{i=1}^{m} \mathbb{E}(Y_i),$$

where $(b)$ follows from Markov inequality. ∎

**Lemma 2.3** *Let $k \geq 1$. Then $\Delta^k V_j(x) \leq -k\epsilon_j$, $1 \leq j \leq J$, for $x \in \mathcal{A}_{j,k}^c$.*

*Proof:* Let $x \in \mathcal{A}_{j,k}^c$. Then $\Delta^k V_j(x)$

$$= \sum_{y \in \mathcal{X}} \Delta V_j(y) \sum_{l=0}^{k-1} p_{xy}^l$$

$$= \sum_{y \in \mathcal{X}_j} \Delta V_j(y) \sum_{l=0}^{k-1} p_{xy}^l + \sum_{y \in \mathcal{X}_j^c} \Delta V_j(y) \sum_{l=0}^{k-1} p_{xy}^l$$

$$\overset{(a)}{\leq} -k\epsilon_j$$

where $(a)$ follows from Assumption 2.2. ∎

Let us assume that $\Delta V_j(x)$ assumes $L_j$ different values $d_{j,1} > d_{j,2} > \cdots > d_{j,L_j}$ on the set $\mathcal{X}$. For $1 \leq k \leq L_j$, define $A_{j,k} = \{x \in \mathcal{X} : \Delta V_j(x) = d_{j,k}\}$. It is obvious to note that the collection of sets $\{A_{j,k}, 1 \leq k \leq L_j\}$ defines a partition of the set $\mathcal{X}$. We now deduce that, for $1 \leq l \leq L_j$, the set $\bigcup_{k=l}^{L_j} A_{j,k}$ is an Upper Set. This simple fact follows from the Assumption 2.3 of Theorem 2.1 that $\Delta V_j(x)$ is a non-increasing function with respect to the partial order $\preceq$. Alternately, the set $\bigcup_{k=1}^{l} A_{j,k}$ is a Lower Set. Next, for $1 \leq j \leq J$, let us define the collection $\{B_{j,k}, 1 \leq k \leq L_j\}$ of Lower sets of $\mathcal{X}$ as

$$B_{j,k} = \{x \in \mathcal{X} : \Delta V_j(x) \geq d_{j,k}\} = \bigcup_{l=1}^{k} A_{j,l}$$

Now, we express $\Delta^n V_j(x)$ as a *weighted sum of probabilities of Lower sets* in $\mathcal{X}$.

**Lemma 2.4**

$$\Delta^n V_j(x) = d_{j,L_j} + \sum_{l=1}^{L_j-1} (d_{j,l} - d_{j,l+1}) \sum_{k=0}^{n-1} p_{x,B_{j,l}}^k, \qquad 1 \leq j \leq J$$

∎

*Proof:*

$$\Delta^n V_j(x) = \sum_{y \in \mathcal{X}} \Delta V_j(y) \sum_{k=0}^{n-1} p_{x,y}^k$$

$$= \sum_{l=1}^{L_j} \sum_{y \in A_{j,l}} \Delta V_j(y) \sum_{k=0}^{n-1} p_{x,y}^k$$

$$= \sum_{l=1}^{L_j} d_{j,l} \sum_{k=0}^{n-1} p_{x,A_{j,l}}^k$$

$$= d_{j,L_j} + \sum_{l=1}^{L_j-1} (d_{j,l} - d_{j,L_j}) \sum_{k=0}^{n-1} p_{x,A_{j,l}}^k$$

$$= d_{j,L_j} + (d_{j,L_j-1} - d_{j,L_j}) \sum_{k=0}^{n-1} p_{x,B_{j,L_j-1}}^k + \sum_{l=1}^{L_j-2} (d_{j,l} - d_{j,L_j-1}) \sum_{k=0}^{n-1} p_{x,A_{j,l}}^k$$

Continuing this way, finally, we obtain the expression stated in Lemma 2.4.

∎

**Lemma 2.5** *The sequences $\{\frac{1}{n}\Delta^n V_j(x), n \geq 1\}$, indexed by the elements $x \in \mathcal{X}$, are uniformly upper bounded, i.e. for a given $\delta > 0$ there exists a $N(\delta)$ such that $\frac{1}{n}\Delta^n V_j(x) \leq c_j^* + \delta$ for all $x \in \mathcal{X}$ and $n \geq N(\delta)$.* ∎

*Proof:* Because the stochastic recursive relation $f$ (equation (1)) is assumed to be order-preserving, the Strassen's Theorem 2.2 implies that for any two states $x$ and $y$ such that $x \preceq y$ it is true that $(X^n|X^0 = x) \leq_{\text{st}} (X^n|X^0 = y)$. In particular, we have that $p((X^n|X^0 = x) \in \mathcal{B}) \geq p((X^n|X^0 = y) \in \mathcal{B})$ for any Lower set $\mathcal{B} \subseteq \mathcal{X}$. But this observation in conjunction with Lemma 2.4 allows us to deduce that $\Delta^n V_j(x) \geq \Delta^n V_j(y)$ whenever $x \preceq y$. In particular, $\Delta^n V_j(y) \leq \Delta^n V_j(x^*)$ where $x^*$ is the minimal element of $\mathcal{X}$. Now, for a $\delta > 0$, we can find $N(\delta)$ such that $\frac{1}{n}\Delta^n V_j(x^*) \leq c_j^* + \delta$. This completes the proof. ∎

Let us fix an arbitrary $\delta > 0$. From Lemma 2.5, it follows that there exits a positive integer $\mathsf{K}$ such that $\frac{\Delta^k V_j(x)}{k} \leq c_j^* + \delta$ for $k \geq \mathsf{K}$ and $1 \leq j \leq J$. Let us pick one such $\mathsf{K}$, and then introduce the set of functions $\{g_j^{\mathsf{K}}(x); x \in \mathcal{X}\}$ such that the following holds:

$$\Delta^{\mathsf{K}} V_j(x) = -g_j^{\mathsf{K}}(x) + \mathsf{K}\left(c_j^* + \delta\right) \tag{4}$$

Two observations on the functions $g_j^{\mathsf{K}}$ are in order: the first and the obvious observation is that $g_j^{\mathsf{K}}(x) \geq 0$ for $x \in \mathcal{X}$. Also, since $\Delta^{\mathsf{K}} V_j(x) \leq -\mathsf{K}\epsilon_j$ for $x \in \mathcal{A}_{j,\mathsf{K}}^c$, we have that $g_j^{\mathsf{K}}(x) \geq \left(c_j^* + \delta + \epsilon_j\right)$ for $x \in \mathcal{A}_{j,\mathsf{K}}^c$. Set $\epsilon = \min_j \epsilon_j$ and $\delta = \min_j \delta_j$. As a result, we have the obvious deduction that $\max_j g_j^{\mathsf{K}}(x) \geq \min_j \mathsf{K}\left(c_j^* + \delta + \epsilon_j\right) = (c^* + \delta + \epsilon)$ for $x \in \cup_j \mathcal{A}_{j,\mathsf{K}}^c$. Hence $\max_j g_j^{\mathsf{K}}(x) < \min_j \mathsf{K}(c^* + \delta + \epsilon)$ implies that $x \in \cap_j \mathcal{A}_{j,\mathsf{K}}$. We should note that $x \in \cap_j \mathcal{A}_{j,k}$ need not imply that $\max_j g_j^{\mathsf{K}}(x) < \mathsf{K}(c^* + \delta + \epsilon)$.

## 2.2 Proof of Theorem 2.1

*Proof:* Denote by $\mathsf{E}_x\left(g_j^{\mathsf{K}}(X^n)\right)$ the expectation of $g_j^{\mathsf{K}}(X^n)$ given that $X^0 = x$ and by $p_x(X^n \in A)$ the probability that $X^n \in A$ given that $X^0 = x$. Now

$$\begin{aligned}
\frac{\Delta^{n\mathsf{K}} V_j(x)}{n} &= \sum_{y \in \mathcal{X}} \frac{1}{n} \sum_{l=0}^{n-1} p_{xy}^{l\mathsf{K}} \Delta^{\mathsf{K}} V_j(y) \\
&\stackrel{(a)}{=} \sum_{y \in \mathcal{X}} \frac{1}{n} \sum_{l=0}^{n-1} p_{xy}^{l\mathsf{K}} \left[-g_j^{\mathsf{K}}(y) + \mathsf{K}(c_j^* + \delta)\right] \\
&= -\sum_{y \in \mathcal{X}} \frac{1}{n} \sum_{l=0}^{n-1} p_{xy}^{l\mathsf{K}} g_j^{\mathsf{K}}(y) + \mathsf{K}(c_j^* + \delta) \\
&= -\frac{1}{n} \sum_{l=0}^{n-1} \mathsf{E}_x\left(g_j^{\mathsf{K}}\left(X^{l\mathsf{K}}\right)\right) + \mathsf{K}(c_j^* + \delta)
\end{aligned}$$

where $(a)$ follows from (4).

Since $\lim_{n \to \infty} \frac{\Delta^{n\mathsf{K}} V_j(x)}{n\mathsf{K}} = c_j^*$, we have that $\frac{1}{n} \sum_{l=0}^{n-1} \mathsf{E}_x\left(g_j^{\mathsf{K}}\left(X^{l\mathsf{K}}\right)\right) = \mathsf{K}\delta$. Now

$$\liminf_{n\to\infty} \frac{1}{n} p_x \left( \max_j g_j^{\mathsf{K}}(x) < \mathsf{K}(c^* + \delta + \epsilon) \right) \overset{(b)}{\geq} 1 - \limsup_{n\to\infty} \frac{1}{n} \sum_{l=0}^{n-1} \sum_{j=1}^{J} \frac{\mathsf{E}_x \left( g_j^{\mathsf{K}}(X^{l\mathsf{K}}) \right)}{\mathsf{K}(c^* + \delta + \epsilon)}$$

$$\geq 1 - \frac{1}{\mathsf{K}(c^* + \delta + \epsilon)} \times$$

$$\sum_{j=1}^{J} \limsup_{n\to\infty} \frac{1}{n} \sum_{l=0}^{n-1} \mathsf{E}_x \left( g_j^{\mathsf{K}}(X^{l\mathsf{K}}) \right)$$

$$= 1 - \frac{J\mathsf{K}\delta}{\mathsf{K}(c^* + \delta + \epsilon)}$$

$$= 1 - \frac{J\delta}{(c^* + \delta + \epsilon)},$$

where $(b)$ follows from Lemma 2.2.

We note that there exists a $\delta_0 > 0$ such that $1 - \frac{J\delta}{(c^* + \delta_0 + \epsilon)} > 0$. Define the set

$$\mathcal{A}_0 = \left\{ x \in \mathcal{X} : \max_j g_j^{\mathsf{K}}(x) < \mathsf{K}(c^* + \delta_0 + \epsilon) \right\}$$

We can observe that $\mathcal{A}_0 \subseteq \cap_j \mathcal{A}_{j,k}$ is a finite set. Hence it follows that for the finite set $\mathcal{A}_0$,

$$\liminf_{n\to\infty} \frac{1}{n} \sum_{l=0}^{n-1} p_x \left( X^{l\mathsf{K}} \in \mathcal{A}_0 \right) > 0$$

Since the chain is assumed to be irreducible and aperiodic, it follows that the Markov chain is positive recurrent. ∎

**Remarks:** Rosberg [3] assumed in his model that $\mathcal{X} = \mathbb{Z}_+^J$ for some integer $J \geq 2$, and then considered an equal number $J$ of partitions $\{\mathcal{X}_j, \mathcal{X}_j^c\}$, $1 \leq j \leq J$, of the countable space $\mathcal{X}$, and also the same number $J$ of Lyapunov functions $\{V_j(x), x \in \mathcal{X}\}$, $1 \leq j \leq J$. However, in our generalization of his criterion, we do not require the countable space $\mathcal{X}$ to have a fixed predetermined dimension. This is reflexed in Assumption 2.1 of Theorem 2.1. Hence we are free to choose an appropriate number of Lyapunov functions and the corresponding suitable partitions of the state space $\mathcal{X}$. We believe this generalization will be useful for the reason that in many situations of interest one does not obtain a state space of some fixed predetermined dimension. Moreover, even in the context of a Markov chain with some fixed dimension it may not be appropriate to consider an equal number $J$ of Lyapunov functions and hence an equal number $J$ of partitions of the state space $\mathcal{X}$. Finally, Assumption 2.3 of Theorem 2.1 is a refinement of the following definition 2.1 proposed by Rosberg. But to appreciate this, we need to discuss model considered by Rosberg in some more detail.

Rosberg considered a fixed dimensional non-negative integer space $\mathbb{Z}_+^J$ and then for each dimension $j$, $1 \leq j \leq J$, he assumed that there exist positive integers $N_j$ and the corresponding partitions $\left\{ \mathcal{X}_{j,N_j}, \mathcal{X}_{j,N_j}^c \right\}$ of the state space $\mathcal{X}$ where $\mathcal{X}_{j,N_j}^c = \{x \in \mathcal{X} : x_j \geq N_j\}$ and $\mathcal{X}_{j,N_j} = \{x \in \mathcal{X} : x_j < N_j\}$. Then he assumed that

$$\Delta V_j(x) \leq \begin{cases} \eta_j & \text{for } x \in \mathcal{X} \\ -\epsilon_j & \text{for } x \in \mathcal{X}^c_{j,N_j}, \end{cases} \tag{5}$$

where $\epsilon_j > 0$ and $\eta_j \geq 0$.

He also made another assumption that there exists a positive integer $M$ such that $p_{xy} = 0$ whenever $(x_j - y_j) > M$ for some $j$, where $x, y \in \mathcal{X}$. We can note here that Assumption 2.2 of Theorem 2.1 is a generalization of the assumption made by Rosberg.

**Definition 2.1 (Rosberg [3])** *For $1 \leq j \leq J$, the sequence $\left\{\frac{\Delta^k V_j(x)}{k}, k \geq 1\right\}$ is said to be uniformly upper bounded (UUB) if, for any $\delta > 0$, there exists a positive integer $\mathsf{K}$ such that $\frac{\Delta^k V_j(x)}{k} < c_j^* + \delta$ for $k \geq \mathsf{K}$ and $x \in \mathcal{X}_{j,N_j+(k-1)M}$.*

But now it is straightforward to see that Lemma 2.5 implies UUB property of Rosberg.

## 3 Stability Analysis of Slotted-Aloha Protocol with Finite Number of Queues

The rest of this section is organized as follows. In Section 3.1, we present modeling details of the standard slotted-Aloha protocol. In Section 3.2, we describe in detail a dominant queueing model of the protocol and then present its stability analysis in Section 3.3. In Section 3.4, we make some remarks on the instability of the protocol.

### 3.1 Model

Consider a system $\mathcal{S}_1$ of $J$ transmitting stations. At each station, there is a queue with infinite buffer space to store incoming packets and the queue is connected to a transmitter. These $J$ transmitters wish to send packets in their respective buffers to a common receiver over a *collision channel*. Transmitter $j$ is assumed to be associated with a Bernoulli random process $Y_j = \{Y_j^n, n \geq 1\}$ where the random variable $Y_j^n$ with the distribution $p\left(Y_j^n = 1\right) = p_j = 1 - p\left(Y_j^n = 0\right)$ models the packet transmission attempt of the transmitter $j$ in the $n$th time slot. That is, the $j$th transmitter with non-empty queue transmits in a slot with probability $p_j$ and does not transmit with probability $\bar{p}_j = 1 - p_j$, independent of everything else. We denote by $Y^n$ the random vector $(Y_1^n, Y_2^n, \ldots, Y_J^n)$. The communication channel between the transmitters and the receiver is modeled by a collision channel model. Under the collision channel model, a packet transmission is successful *if and only if* at most one transmitter with a non-empty queue transmits. When more than one transmitter transmits in a slot, all packet transmissions involved in that time slot are considered to have collided and hence are lost for all practical purposes. The length of a time slot is taken to be the duration of a packet transmission. At the end of each time slot, all transmitting stations are provided with ternary feedback which tells whether the time slot was idle (no attempted transmissions), successful (exactly one transmitter transmitted), or a failure (at least two transmitters transmitted in that time slot).

To model packet arrivals into various queues, we assume that packets arrive randomly into various queues and the packet arrival process into queue $j$ is modeled by an i.i.d. batch arrival process $\Lambda_j = \{\Lambda_j^n, n \geq 1\}$ where the random variable $\Lambda_j^n$ with finite first moment $\mathsf{E}\left(\Lambda_j^n\right) = \lambda_j$ models the number of packet arrivals into queue $j$ during the $n$th time slot. Define by $\lambda$ the vector $(\lambda_1, \lambda_2, \ldots, \lambda_J)$ of packet arrival rates. Let $\overline{Q}_j^n$ denote the number of packets present in the queue $j$ at the beginning of the $n$th time slot. Denote by $\overline{Q}^n = \left(\overline{Q}_1^n, \overline{Q}_2^n, \ldots, \overline{Q}_J^n\right)$ the queue-length vector at the beginning of the $n$th time slot. From the assumptions made so far, we can easily note that the queue-length process $\{\overline{Q}^n, n \geq 0\}$ is a discrete-time Markov chain over the countable space $\mathbb{Z}_+^J$, where $\mathbb{Z}_+^J$ is the set of non-negative integer vectors of dimension $J$.

For an event $A$, let us define the indicator function $\mathbb{I}\{A\}$ as $\mathbb{I}\{A\} = 1$ if the event $A$ is true, and $\mathbb{I}\{A\} = 0$ otherwise. From the above discussion, it is clear that transmitter $j$ transmits a packet *if and only if* the product $Y_j^n \mathbb{I}\{\overline{Q}_j^n > 0\} = 1$, and *no* packet transmission happens otherwise. If a packet from the $j$th queue is involved in collision during the $n$th time slot it is then retransmitted in the $(n+1)$th time slot with the same probability $p_j$. When there is only one transmission in a time slot we say that the transmission is successful in that it is received error free at the receiver and the corresponding queue length is decremented by 1. The queue length evolution with time is given by

$$\left.\begin{aligned}\overline{Q}_j^{n+1} &= \overline{Q}_j^n + \Lambda_j^n - \overline{D}_j^n \\ \overline{D}_j^n &= Y_j^n \mathbb{I}\{\overline{Q}_j^n > 0\} \prod_{k \neq j}\left(1 - Y_k^n \mathbb{I}\{\overline{Q}_k^n > 0\}\right)\end{aligned}\right\} \quad (6)$$

where the random variable $\overline{D}_j^n \in \{0, 1\}$ denotes the number of departures from the $j$th queue in the $n$th time slot.

## 3.2 Dominant System

We now consider another system $\mathcal{S}_2$ of $J$ queues such that when $\mathcal{S}_1$ and $\mathcal{S}_2$ have the following *identical* features **F1**, **F2**, and **F3**, then $\mathcal{S}_2$ will *dominate* the original system $\mathcal{S}_1$ in the sense that queue lengths in $\mathcal{S}_2$ will be at least as large as the respective queue lengths in $\mathcal{S}_1$ at all times. Let the random variables $Q_j^n$ and $D_j^n$ denote [1] the queue length of and the number of departures from the $j$th queue for the $n$th time slot. The following features are assumed to be identical to both $\mathcal{S}_1$ and $\mathcal{S}_2$.

**F1** initial state, i.e., $Q^0 = \overline{Q}^0$.

**F2** arrival processes, i.e., arrivals into the $j$th queue in $\mathcal{S}_2$ occur *exactly* at the same time instants as in the original system $\mathcal{S}_1$.

**F3** transmission attempts, i.e., the Bernoulli random vector $Y^n$ that determines transmission attempts in the original system $\mathcal{S}_1$ for the $n$th time slot also determines the transmission attempts for the $n$th time slot in the system $\mathcal{S}_2$.

The distinguishing feature of $\mathcal{S}_2$ from $\mathcal{S}_1$ will come from the presence of *dummy packet transmissions* from $\mathcal{S}_2$, i.e., queue $j$ of $\mathcal{S}_2$ transmits a packet, called dummy packet, with probability $p_j$ upon becoming empty. The aspect on which the two systems will differ is the interference as seen by the individual queues in $\mathcal{S}_2$. By careful construction, we make interference for the queue $j$ in $\mathcal{S}_2$ at least as large as the interference seen by the queue $j$ in the original system $\mathcal{S}_1$. As a consequence, a successful transmission from the queue $j$ in $\mathcal{S}_2$ implies a successful transmission from the queue $j$ of $\mathcal{S}_1$ provided $\overline{Q}_j > 0$. But the converse need not be true. This fact becomes revealed when we compare the queue length evolutions (6) and (7), respectively, of $\mathcal{S}_1$ and $\mathcal{S}_2$. Henceforth, we will refer to $\mathcal{S}_2$ as a *dominant* of the original $\mathcal{S}_1$.

---

[1] Usage of an *over line* in the notation will distinguish queue length and departure random variables of $\mathcal{S}_1$ from that of $\mathcal{S}_2$. An over line in the notation is used *only* for the original system $\mathcal{S}_1$.

To be able to define the rules that will specify the interference to be seen by any individual queue in $\mathcal{S}_2$, we define two sets $\mathcal{U}_j$ and $\mathcal{V}_j$ of queues for each queue $j$.

$$\mathcal{U}_1 = \emptyset \text{ and } \mathcal{U}_j = \{1, 2, \ldots, j-1\} \text{ for } j \geq 2$$
$$\mathcal{V}_j = \{j+1, j+2, \ldots, J\} \text{ for } j < J \text{ and } \mathcal{V}_J = \emptyset$$

The sets $\mathcal{U}_j$ and $\mathcal{V}_j$ will be designated as, respectively, the set of *non-persistent* and the set of *persistent* queues of the $j$th queue for the following reason. A real packet transmission from queue $j$ is effected by (i) *only real* packet transmissions from the queues that belong to the set of queues $\mathcal{U}_j$, and (ii) *both real and dummy* packet transmissions from the queues that belong to the set of queues $\mathcal{V}_j$. In other words, each queue $j$ in $\mathcal{S}_2$ transmits a *dummy* packet with probability $p_j$ upon becoming empty. But not every real packet transmission is effected by a dummy packet transmission from the $j$th queue. Dummy transmissions are designed only to cause interference but the successful transmission of a dummy packet from the queue $j$ has no significance, i.e., queue length $Q_j$ is unaffected. An interesting point and the main aspect in which our dominant system $\mathcal{S}_2$ differs from the previous constructions is that different queues have different sets of persistent and non-persistent queues in the *same time slot*.

From the discussion made above, the queue length evolution in the dominant system $\mathcal{S}_2$ can now be represented as

$$\left. \begin{array}{l} Q_j^{n+1} = Q_j^n + \Lambda_j^n - D_j^n \\ \\ D_j^n = \left[\displaystyle\prod_{k \in \mathcal{U}_j} (1 - Y_k^n \mathbb{I}\{Q_k^n > 0\})\right] Y_j^n \mathbb{I}\{Q_j^n > 0\} \left[\displaystyle\prod_{k \in \mathcal{V}_j} (1 - Y_k^n)\right] \end{array} \right\} \quad (7)$$

For the queue-length vector $Q$, we define $u_j(Q)$ as the probability that *no* real packet is transmitted from the queues of the set $\mathcal{U}_j$. That is

$$u_1(Q) = 1$$
$$u_j(Q) = \prod_{k=1}^{j-1} (1 - p_k \mathbb{I}\{Q_k > 0\}) \quad \text{for } j \geq 2$$

Similarly, $v_j(Q)$ will be defined as the probability that *neither* a real packet transmission *nor* a dummy packet transmission will occur from the queues of the set $\mathcal{V}_j$, i.e.,

$$v_j(Q) = \prod_{k=j+1}^{J} \overline{p}_k \quad \text{for } 1 \leq j \leq J-1$$
$$v_J(Q) = 1$$

We note here that $v_j(Q)$ is queue-length *independent* and $u_j(Q)$ is queue-length *dependent*. Likewise, we define the *success probability* $r_j(Q)$ of the $j$th queue as

$$r_j(Q) = u_j(Q)p_j v_j(Q)\mathbb{I}\{Q_i > 0\}$$

Define the success probability vector $r(Q) = (r_1(Q), r_2(Q), \ldots, r_J(Q))$. For the sake of notational convenience, henceforth, we denote the more expressive notation $u_j(Q)$, $v_j(Q)$, $r_j(Q)$, and $r(Q)$, respectively, as simply $u_j$, $v_j$, $r_j$, and $r$, as long as no ambiguity is caused. Some times we may also write $u_j(r)$ in place of $u_j(Q)$ or $u_j$. From the knowledge of $r$, we can immediately tell which queues are empty and which queues are non-empty and hence indices of the non-empty queues in the $\mathcal{U}_j$. Then it becomes straightforward to write down the value of $u_j(r)$. Also, in the rest of this paper, we will prefer the more convenient notation $Q_j$ in place of $Q_j^n$ unless explicit emphasis on the time slot is needed, and we extend this rule to other variables too. With the help of the notation introduced so far, we now state our central result on stability of the slotted-Aloha protocol.

**Proposition 3.1** *Let $\eta = (\eta_1, \eta_2, \ldots, \eta_J)$ denote a permutation of the set $\{1, 2, \ldots, J\}$. Define $\mathcal{C}(\eta) \subset \mathbb{R}_+^J$ to be the set of $\lambda$ that satisfies*

$$\frac{\lambda_{\eta_j}}{p_{\eta_j} v_{\eta_j}} + \mathbb{I}\{j \geq 2\} \sum_{k=1}^{j-1} \frac{\lambda_{\eta_k}}{v_{\eta_k}} < 1, \quad \text{for } 1 \leq j \leq J$$

*Define $\mathcal{C} = \cup_\eta \mathcal{C}(\eta)$. Then the dominant system $\mathcal{S}_2$ is stable for $\lambda \in \mathcal{C}$.* ∎

Since stability of $\mathcal{S}_2$ implies stability of $\mathcal{S}_1$ because of queue length dominance, we conclude that the original system $\mathcal{S}_1$ is also stable for $\lambda$ that satisfies the conditions of Proposition 3.1. In Figure 1, we show a portion of the stability region $\mathcal{C}$ when $J = 3$.

We will now specialize Proposition 3.1 for $J = 2, 3$, and $J \to \infty$. For $J = 2$, we have

$$\mathcal{C}(\{1,2\}) = \left\{\lambda : \frac{\lambda_1}{p_1 \bar{p}_2} < 1 \text{ and } \frac{\lambda_1}{\bar{p}_2} + \frac{\lambda_2}{p_2} < 1\right\}$$

$$\mathcal{C}(\{2,1\}) = \left\{\lambda : \frac{\lambda_2}{\bar{p}_1 p_2} < 1 \text{ and } \frac{\lambda_1}{p_1} + \frac{\lambda_2}{\bar{p}_1} < 1\right\}$$

Then $\mathcal{C} = \mathcal{C}(\{1,2\}) \cup \mathcal{C}(\{2,1\})$ reduces to the exact stability condition derived by Tsybakov and Mikhailov [2].

Let us now consider $J = 3$. We have a total of six permutations of the set of queues $\{1, 2, 3\}$ and the corresponding sufficient conditions for stability are as follows:

$$\mathcal{C}(\{1,2,3\}) = \left\{\lambda : \begin{array}{l} \dfrac{\lambda_1}{p_1 \bar{p}_2 \bar{p}_3} < 1 \\[6pt] \dfrac{\lambda_2}{p_2 \bar{p}_3} + \dfrac{\lambda_1}{\bar{p}_2 \bar{p}_3} < 1 \\[6pt] \dfrac{\lambda_3}{p_3} + \dfrac{\lambda_2}{\bar{p}_3} + \dfrac{\lambda_1}{\bar{p}_2 \bar{p}_3} < 1 \end{array}\right\}$$

$$\mathcal{C}(\{1,3,2\}) = \left\{\lambda : \begin{array}{c} \dfrac{\lambda_1}{p_1\bar{p}_2\bar{p}_3} < 1 \\ \dfrac{\lambda_3}{\bar{p}_2 p_3} + \dfrac{\lambda_1}{\bar{p}_2\bar{p}_3} < 1 \\ \dfrac{\lambda_2}{p_2} + \dfrac{\lambda_3}{\bar{p}_2} + \dfrac{\lambda_1}{\bar{p}_2\bar{p}_3} < 1 \end{array}\right\}$$

$$\mathcal{C}(\{2,3,1\}) = \left\{\lambda : \begin{array}{c} \dfrac{\lambda_2}{p_2\bar{p}_1\bar{p}_3} < 1 \\ \dfrac{\lambda_3}{p_3\bar{p}_1} + \dfrac{\lambda_2}{\bar{p}_1\bar{p}_3} < 1 \\ \dfrac{\lambda_1}{p_1} + \dfrac{\lambda_3}{\bar{p}_1} + \dfrac{\lambda_2}{\bar{p}_1\bar{p}_3} < 1 \end{array}\right\}$$

$$\mathcal{C}(\{2,1,3\}) = \left\{\lambda : \begin{array}{c} \dfrac{\lambda_2}{p_2\bar{p}_1\bar{p}_3} < 1 \\ \dfrac{\lambda_1}{p_1\bar{p}_3} + \dfrac{\lambda_2}{\bar{p}_1\bar{p}_3} < 1 \\ \dfrac{\lambda_3}{p_3} + \dfrac{\lambda_1}{\bar{p}_3} + \dfrac{\lambda_2}{\bar{p}_1\bar{p}_3} < 1 \end{array}\right\}$$

$$\mathcal{C}(\{3,2,1\}) = \left\{\lambda : \begin{array}{c} \dfrac{\lambda_3}{p_3\bar{p}_1\bar{p}_2} < 1 \\ \dfrac{\lambda_2}{p_2\bar{p}_1} + \dfrac{\lambda_3}{\bar{p}_1\bar{p}_2} < 1 \\ \dfrac{\lambda_1}{p_1} + \dfrac{\lambda_2}{\bar{p}_1} + \dfrac{\lambda_3}{\bar{p}_1\bar{p}_2} < 1 \end{array}\right\}$$

$$\mathcal{C}(\{3,1,2\}) = \left\{\lambda : \begin{array}{c} \dfrac{\lambda_3}{p_3\bar{p}_1\bar{p}_2} < 1 \\ \dfrac{\lambda_1}{p_1\bar{p}_2} + \dfrac{\lambda_3}{\bar{p}_1\bar{p}_2} < 1 \\ \dfrac{\lambda_2}{p_2} + \dfrac{\lambda_1}{\bar{p}_2} + \dfrac{\lambda_3}{\bar{p}_1\bar{p}_2} < 1 \end{array}\right\}$$

Each of these six sufficient conditions for $J = 3$ strictly include the respective sufficient conditions derived in Rao and Ephremides [4]. For the asymptotic case of $J \to \infty$ and symmetric arrival rates and transmission probabilities (i.e., $\lambda_j = \lambda$ and $p_j = p$), our result recovers the well known result [12] that the system is unstable.

## 3.3 Positive Recurrence of the Queue Length Process $\{Q^n, n \geq 1\}$

We now prove Proposition 3.1 for the particular permutation $\eta = (1, 2, \ldots, J)$. The proof consists of verifying validity of the Assumptions 2.1, 2.2, and 2.3 of Theorem 2.1 for the Markov chain $\{Q^n, n \geq 1\}$. Then Theorem 2.1 implies the sufficiency condition of Proposition 3.1. But to facilitate the presentation, we need to introduce some more notation. We say that $Q$ and $Q'$ are "component-wise" partially ordered and write $Q \preceq Q'$ if $Q_j \leq Q'_j$ for $1 \leq j \leq J$.

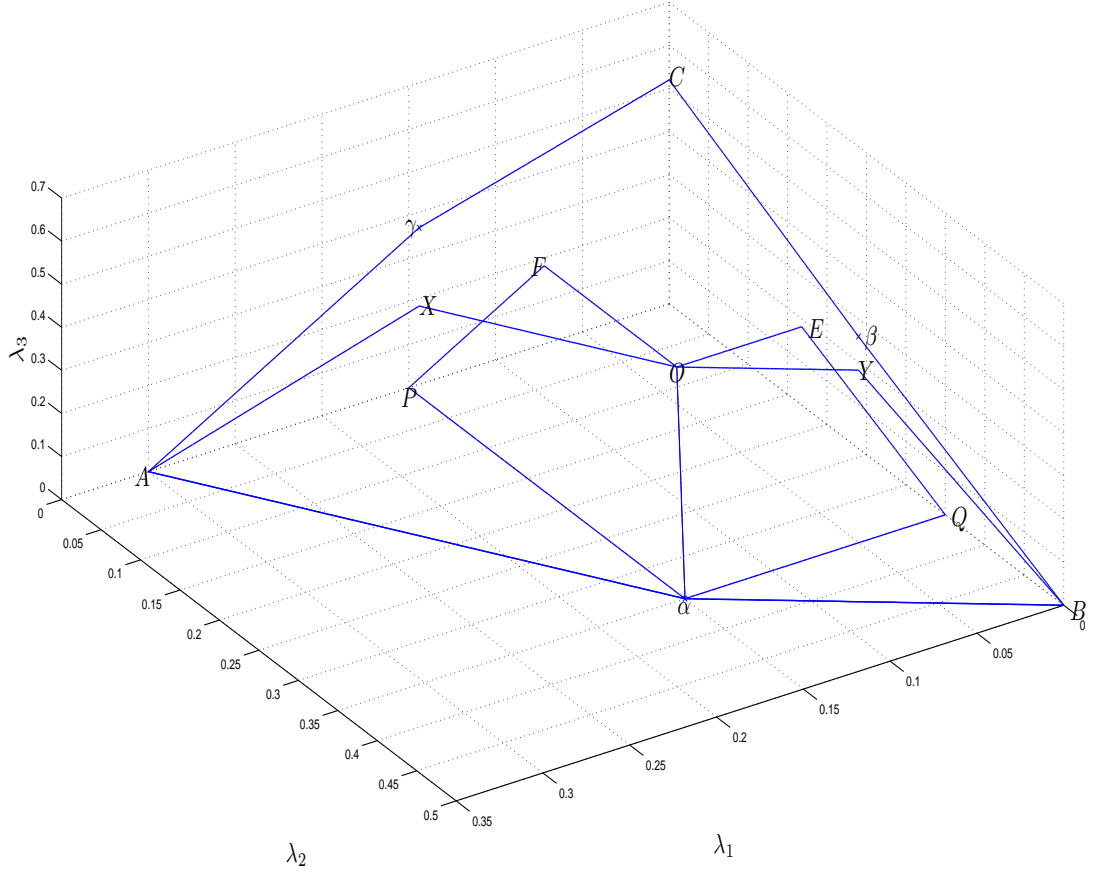

Figure 1: Figure shows the sufficiency conditions for stability for the permutations $\{3, 2, 1\}$ and $\{3, 1, 2\}$. In the figure, $A = (p_1, 0, 0)$, $B = (0, p_2, 0)$, $C = (0, 0, p_3)$, $O = (p_1\bar{p}_2\bar{p}_3, \bar{p}_1 p_2 \bar{p}_3, \bar{p}_1 \bar{p}_2 p_3)$, $\alpha = (p_1 \bar{p}_2, \bar{p}_1 p_2, 0)$, $\beta = (0, p_2 \bar{p}_3, \bar{p}_2 p_3)$, $\gamma = (p_1 \bar{p}_3, 0, \bar{p}_1 p_3)$, $P = (p_1 \bar{p}_2, 0, 0)$, $Q = (0, \bar{p}_1 p_2, 0)$, $X = (p_1 \bar{p}_3, 0, \bar{p}_1 \bar{p}_2 p_3)$, $Y = (0, p_2 \bar{p}_3, \bar{p}_1 \bar{p}_2 p_3)$, $E = (0, \bar{p}_1 p_2 \bar{p}_3, \bar{p}_1 \bar{p}_2 p_3)$, $F = (p_1 \bar{p}_2 \bar{p}_3, 0, \bar{p}_1 \bar{p}_2 p_3)$. When $\lambda_3 = 0$, the stability region is confined to $\lambda_1 \lambda_2$-plane and the corresponding stability region boundary is described by the line segments $A\alpha$ and $B\alpha$. Similarly, the line segments $A\gamma$ and $C\gamma$ together represent the boundary of the stability region when $\lambda_2 = 0$, and the line segments $B\beta$ and $C\beta$ together represent the boundary of the stability region when $\lambda_1 = 0$. For $\lambda_3 < \bar{p}_1 \bar{p}_2 p_3$, the plane segment $A\alpha OX$, bearing the plane equation $\frac{\lambda_1}{p_1} + \frac{\lambda_2}{\bar{p}_1} + \frac{\lambda_3}{\bar{p}_1 \bar{p}_2} = 1$ and the plane segment $Q\alpha OE$, bearing the plane equation $\frac{\lambda_2}{\bar{p}_1 p_2} + \frac{\lambda_3}{\bar{p}_1 \bar{p}_2} = 1$, together describe the sufficiency condition for stability.

We note that the distribution of the departure random variable $D_j$ is $p(D_j = 1) = r_j = 1 - p(D_j = 0)$. Hence follows that

$$\mathsf{E}(D_j) = r_j \qquad (8)$$

Define $\mathcal{R}$ to be the set of all success probability vectors, i.e. $\mathcal{R} = \{r(Q) : Q \in \mathbb{Z}_+^J\}$. For $1 \leq j \leq J$, we define a partition $\{\mathcal{R}_j, \mathcal{R}_j^c\}$ of the set $\mathcal{R}$ as

$$\mathcal{R}_j = \{r \in \mathcal{R} : r_j = 0\} \quad \text{and} \quad \mathcal{R}_j^c = \{r \in \mathcal{R} : r_j > 0\}$$

Define the mapping $g : \mathbb{Z}_+^J \to \mathcal{R}$ by $g(Q) = (r_1(Q), r_2(Q), \ldots, r_J(Q))$,

where $g(Q)$ is the vector of success probabilities when the queue length vector is $Q$. Since the knowledge of which queues are empty and which queues are non-empty alone is sufficient to determine the success probability vector $r$, we can group all queue length vectors $Q$ that result in the same $r$. This is done by defining the *set-valued* map $g^{-1} : \mathcal{R} \to \mathbb{Z}_+^J$ as

$$g^{-1}(r) \triangleq \{Q \in \mathbb{Z}_+^J : g(Q) = r\}$$

We note that the collection $\{g^{-1}(r), r \in \mathcal{R}\}$ of sets defines a partition of the space queue length vectors, $\mathbb{Z}_+^J$.

Next, we prove that there exist positive weights such that *the sum of weighted expected number of departures from the queues that belong to the set* $\{1, 2, \ldots, j\}$ conditioned on the event $\{Q_j \geq 1\}$ equals one, and when conditioned on the event $\{Q_j = 0\}$, equals the probability $(1 - u_j)$ that there is at least one transmission from the set of queues $\mathcal{U}_j$. We establish this fact for every $j$. For brevity, let us define the random variable

$$\hat{D}_j = \frac{D_j}{p_j v_j} + \mathbb{I}\{j \geq 2\} \sum_{k=1}^{j-1} \frac{D_k}{v_k}, \qquad 1 \leq j \leq J$$

**Lemma 3.1**

$$\mathsf{E}\left(\hat{D}_j \,\Big|\, A\right) = \begin{cases} 1 - u_j & \text{if } A = \{Q_j = 0\} \\ 1 & \text{if } A = \{Q_j \geq 1\} \end{cases}$$

∎

*Proof:* Suppose that for some $k_1$ and $k_2$ such that $1 \leq k_1 < k_2 \leq j \leq J$, let it be true that (i) $Q_{k_1} > 0$ and $Q_{k_2} > 0$, and (ii) every other queue $l$, $k_1 < l < k_2$, is such that $Q_l = 0$. Then it is easy to note that $u_{k_2} = \bar{p}_{k_1} u_{k_1}$. As a consequence, we can immediately note that $u_{k_2} + p_{k_1} u_{k_1} = u_{k_1}$. Then

$$\mathsf{E}\left(\hat{D}_j \,\Big|\, Q_j \geq 1\right) \stackrel{(a)}{=} u_j + \mathbb{I}\{j \geq 2\} \sum_{k=1}^{j-1} p_k u_k \mathbb{I}\{Q_k > 0\} \stackrel{(b)}{=} 1$$

where $(a)$ follows from (8) and $(b)$ follows by repeatedly applying the observation made above. Almost on the similar lines, we can also note that

$$\mathsf{E}\left(\hat{D}_j\middle|Q_j=0\right) = \sum_{k=1}^{j-1} p_k u_k \mathbb{I}\{Q_k > 0\} = 1 - u_j$$

∎

Now, we are at a stage to verify the Assumption 2.1. To verify the Assumption 2.1, consider the Lyapunov functions $V_j$, $1 \leq j \leq J$, defined as

$$V_j(Q) = \frac{Q_j}{v_j p_j} + \mathbb{I}\{j \geq 2\} \sum_{k=1}^{j-1} \frac{Q_k}{v_k} \tag{9}$$

Over the state space $\{Q_j \geq 1\}$, the drift $\Delta V_j(Q)$ can be written as

$$\Delta V_j(Q) = \frac{\lambda_j}{v_j p_j} + \mathbb{I}\{j \geq 2\} \sum_{k=1}^{j-1} \frac{\lambda_k}{v_k} - \mathsf{E}\left(\hat{D}_j\middle|Q_j \geq 1\right)$$

$$= \frac{\lambda_j}{v_j p_j} + \mathbb{I}\{j \geq 2\} \sum_{k=1}^{j-1} \frac{\lambda_k}{v_k} - 1 \tag{10}$$

Likewise, the drift $\Delta V_j(Q)$ over the state space $\{Q_j = 0\}$ can be written as

$$\Delta V_j(Q) = \frac{\lambda_j}{v_j p_j} + \mathbb{I}\{j \geq 2\} \sum_{k=1}^{j-1} \frac{\lambda_k}{v_k} - (1 - u_j) \tag{11}$$

where (10) and (11) follow from Lemma 3.1.

We now verify Assumption 2.2. Define $\mathcal{A}_{j,k}^c = \{Q \in \mathbb{Z}_+^J : Q_j \geq k\}$ and $\mathcal{A}_{j,k} = \{Q \in \mathbb{Z}_+^J : Q_j < k\}$. We can easily see that $\cap_{j=1}^J \mathcal{A}_{j,k}$ is a finite set.

We can now make an important observation that with respect to the component-wise partial order the drift $\Delta V_j(Q)$ is a non-increasing function of the argument $Q$. To see this, we note that $\mathsf{E}\left(\hat{D}_j\right)$ is at most one and this value is attained when $Q_j \geq 1$ irrespective of the size of other queues. When $Q_j = 0$, $\mathsf{E}\left(\hat{D}_j\right)$ equals the probability $1 - u_j$ that there is at least one transmission from the set $\mathcal{U}_j$ of queues. But for two states $Q$ and $Q'$ such that $Q_j = Q'_j = 0$ and $Q \leq Q'$, the set of non-empty queues associated with the queue length vector $Q$ is a subset of the set of non-empty queues associated with the queue length vector $Q'$. Hence the the probability that there is at least one transmission from the set $\mathcal{U}_j$ of queues associated with the state $Q'$ is *at least as large* as the corresponding probability associated with the state $Q$. In other words, for $1 \leq j \leq J$, $\mathsf{E}\left(\hat{D}_j\right)$ is a non-decreasing function of the state $Q$, and hence the drift $\Delta V_j(Q)$ is a non-increasing function of the argument $Q$. These arguments essentially verify the second part the Assumption 2.3.

We now prove that the stochastic recursive relation (7) that models the queue lengths evolution of the dominant system $\mathcal{S}_2$ is order preserving. Consider two queue length vectors $Q$ and $Q'$ such that $Q \leq Q'$. Then let us imagine two dominant systems of which the first one is started in state $Q$ and the second one is started in state $Q'$ but both are fed by the identical input processes $\{Y^n, n \geq 1\}$ and $\{\Lambda^n, n \geq 1\}$. An obvious inference is that the set of non-empty queues of $\mathcal{U}_j(Q)$ is contained in the set of non-empty queues of $\mathcal{U}_j(Q')$. Hence a successful transmission from queue $j$ corresponding to the state $Q$ need not imply a successful transmission from queue $j$ of the state $Q'$. In other words, the queue lengths in the dominant system corresponding to the initial state $Q'$ dominate the corresponding queue lengths in the dominant system corresponding to the initial state $Q$, and hence the queue length evolution (7) is order preserving.

## 3.4 Remarks on the Sufficient Conditions for instability of $\mathcal{S}_1$

We use the *indistinguishability* argument of [5] and [4] to derive sufficient conditions for transience of the Markov chain. We shall use a coupling argument to show that with positive probability the dominant system $\mathcal{S}_2$ and the original system $\mathcal{S}_1$ are indistinguishable under the stated sufficient condition for transience of $\mathcal{S}_2$ in Proposition 3.2. Hence instability of the dominant system $\mathcal{S}_2$ also implies instability of the original system $\mathcal{S}_1$. But before we proceed further in this section, we state a sufficient condition for transience in the context of Markov chains over a countable state space.

**Theorem 3.1 (Theorem 8.0.2 (i) of [13])** *An irreducible Markov chain over a countable space $\mathcal{X}$ is transient iff there exists a bounded non-negative function, $V$, and a non-empty set $C \subset \mathcal{X}$ such that for all $x \in C^c$, $\Delta V(x) \geq 0$, and $\exists x \in C^c$ such that $V(x) > \sup_{y \in C} V(y)$.*

**Proposition 3.2** *Let $\eta = (\eta_1, \eta_2, \ldots, \eta_J)$ denote a permutation of the set $\{1, 2, \ldots, J\}$. Define $\mathcal{D}(\eta) \subset \mathbb{R}_+^J$ to be the set of $\lambda$ that satisfies*

$$\frac{\lambda_{\eta_1}}{p_{\eta_1} v_{\eta_1}} < 1$$

$$\frac{\lambda_{\eta_j}}{p_{\eta_j} v_{\eta_j}} + \sum_{k=1}^{j-1} \frac{\lambda_{\eta_k}}{v_{\eta_k}} > 1, \quad \text{for } 2 \leq j \leq J$$

*Define $\mathcal{D} = \cup_\eta \mathcal{D}(\eta)$. Then the dominant system $\mathcal{S}_2$ is unstable for $\lambda \in \mathcal{D}$.* ∎

*Proof:* Let $0 < \theta < 1$. For each $j$, $2 \leq j \leq J$, consider the Lyapunov function $Z_j(Q, \theta) \triangleq 1 - \theta^{V_j(Q)}$, where $V_j(Q)$ is as defined in equation (9). First, we note that the Lyapunov function $Z_j(Q, \theta)$ is bounded. Define the drift $\Delta Z_j(Q, \theta)$ as

$$\Delta Z_j(Q, \theta) = \sum_{Q' \in \mathbb{Z}_+^J} \left[\left(1 - \theta^{V_j(Q')}\right) - \left(1 - \theta^{V_j(Q)}\right)\right] p_{QQ'}$$

$$= \sum_{Q' \in \mathbb{Z}_+^J} \left(\theta^{V_j(Q)} - \theta^{V_j(Q')}\right) p_{QQ'}$$

Differentiating $\Delta Z_j(Q, \theta)$ with respect to $\theta$, we obtain

$$\frac{d\Delta Z_j(Q, \theta)}{d\theta} = \sum_{Q' \in \mathbb{Z}_+^J} \left(V_j(Q)\theta^{V_j(Q)-1} - V_j(Q')\theta^{V_j(Q')-1}\right) p_{QQ'}$$

The following two observations are immediate: (i) $\Delta Z_j(Q,1) = 0$ and (ii) $\frac{d\Delta Z_j(Q,1)}{d\theta} = \sum_{Q' \in \mathbb{Z}_+^J} (V_j(Q) - V_j(Q')) p_{QQ'} = -\Delta V_j(Q)$ where $\Delta V_j(Q)$ is as defined in equations (10) and (11).

Let us now suppose that $\Delta V_j(Q) \geq 0$ over the *entire* state space $\mathbb{Z}_+^J$. Because $\Delta Z_j(Q,\theta)$ is a differentiable function of $\theta$ and also because of the observation (i) and (ii) made above, there exists a $0 < \theta^* < 1$ such that $\Delta Z_j(Q, \theta^*) > 0$ over the state space $\mathbb{Z}_+^J$. Hence the sufficiency condition stated in Theorem 3.1 for transience of the Markov chain $\{Q^n, n \geq 1\}$ is satisfied. Finally, we note that $\frac{\lambda_{\eta_j}}{p_{\eta_j} v_{\eta_j}} + \sum_{k=1}^{j-1} \frac{\lambda_{\eta_k}}{v_{\eta_k}} > 1$ for $2 \leq j \leq J$ imply $\Delta V_j(Q) > 0$ for $2 \leq j \leq J$. ∎

**Remarks:**

1. The indistinguishability argument is based on the fact that the dominant and the original are identical as long as the their queues do not empty.

2. If our aim is to derive sufficient conditions for instability of the dominant system $\mathcal{S}_2$, then the conditions stated in the Proposition 3.2 are *too strong*. The requirement that $\frac{\lambda_j}{v_j p_j} + \mathbb{I}\{j \geq 2\} \sum_{k=1}^{j-1} \frac{\lambda_k}{v_k} > 1$ for *at least* one $j$, $1 \leq j \leq J$, is sufficient for instability of $\mathcal{S}_2$. But we require the conditions under which the dominant system $\mathcal{S}_2$ is *indistinguishable* from the original system $\mathcal{S}_1$ and hence the need for the requirement that $\frac{\lambda_j}{v_j p_j} + \mathbb{I}\{j \geq 2\} \sum_{k=1}^{j-1} \frac{\lambda_k}{v_k} > 1$ for $2 \leq j \leq J$.

# 4 Conclusion

In this paper, we have revisited the stability analysis of slotted-Aloha protocol with finite number of queues by applying Theorem 2.1 which is a generalization of the positive recurrence criterion due to Rosberg [3]. An aim in this paper has been to illustrate how stochastic monotonicity arguments in conjunction with Lyapunov-drift properties can be used in establishing positive recurrence of a Markov chain in a countable space setting. We have seen that two steps are involved in verifying Theorem 2.1. The first step involves verifying Assumptions 2.1 and 2.2. The second step is about verifying Assumption 2.3 and is also the stage where we invoke stochastic monotonicity arguments of the underlying Markov chain. Our experience so far has been that one of these two steps is hard to verify, if not both, depending on the problem. A simplifying feature of this positive recurrence criterion we believe is that it allows one to think of Lyapunov-drift properties confined to certain *proper* subsets of the state space, which is relatively simpler, rather than the entire state space, which is harder. An interesting topic for further research on this problem would be to provide tighter sufficiency conditions for transience of the protocol.